\documentclass{elsarticle}
\usepackage{graphicx}
\usepackage{epstopdf} %converting to PDF
\usepackage{amssymb}
\usepackage{amsmath}
\usepackage{amsfonts}
\usepackage{comment}
\usepackage{subfigure}

\begin{document}

\begin{frontmatter}

\title{Community structures in allelopathic interaction networks: an eco-evolutionary approach}

\author[dpf]{Sylvestre Aureliano Carvalho}
\author[dpf,nistcs]{Marcelo Lobato Martins\corref{cor}}
\ead{mmartins@ufv.br}

\address[dpf]{Departamento de F\'{\i}sica, Universidade Federal de Vi\c{c}osa, 36570-000, Vi\c{c}osa, MG, Brazil}
\address[nistcs]{National Institute of Science and technology for Complex Systems, Centro Brasileiro de Pesquisas F\'{\i}sicas, 
Rua Xavier Sigaud 150, 22290-180, Rio de Janeiro, Brazil}

\cortext[cor]{Corresponding author.}

\date{\today}

\begin{abstract}
Nowadays, evidence is mounting that the race of living organisms for adaptation to the chemicals synthesized by their neighbours may drive community structures. Particularly, some bacterial infections and plant invasions disruptive of the native community rely on the release of allelochemicals that inhibit or kill sensitive strains or individuals from their own or other species. In this report, an eco-evolutionary model for community assembly through resource competition, allelophatic interactions, and evolutionary branching is presented and studied by numerical analysis. Our major findings are that stable communities with increasing biodiversity can emerge at weak allelopathic suppression, but stronger allelophaty is negatively correlated with community diversity. In the former regime the allelopathic interaction networks exhibit Gaussian degree distributions, while in the later one the network degrees are Weibull distributed. 
\end{abstract}


\end{frontmatter}

\section{Introduction}

Conventional explanations of biodiversity postulate that it is passively shaped by niche differentiation, density-dependent predation pressure, habitat heterogeneity, or fluctuations in the resources required by the biological communities. Furthermore, stabilizing mechanisms relying on negative intraspecific interactions, stronger than interspecific interactions, are essential for species coexistence \cite{Chesson} since they cause species to limit themselves more than other organisms. Without stabilizing mechanisms, the inhibitory effects of competition on inferior competitors will ultimately lead to their extinction. Classically, such stabilizing interactions have been thought to result from resource partitioning: competing species can coexist provided they are most limited by different resources and consume the resources they are most limited by at a higher rate than do other species \cite{Tilman}.

However, the astonishing high diversity observed within microorganism communities in seemingly uniform environments --- the famous paradox of the plankton \cite{Hutchinson} --- challenges the conventional resource competition framework. Indeed, even a highly structured habitat can hardly maintain such astronomical species numbers. Moreover, experiments performed with plants have neither shown intraspecific unequivocally exceeding interspecific competition \cite{Goldberg} nor competing plants coexisting through resource partitioning \cite{Miller}. Also, abiotic supply rates seems to be relatively high and stable over time, whereas the resident species do neither reduce resource densities or interfere greatly with resource access \cite{Davis}. 

In contrast, interference competitions mediated by the production of toxic chemical compounds --- antibiotic, phytotoxins, lactate, etc. --- are ubiquitous in biological communities, from microorganisms, such as bacteria \cite{Cordero}, yeasts \cite{Starmer}, and other fungi \cite{Berdy}, to cancer cells \cite{Gatenby,Braganhol} and plant invasions \cite{Bais}. So, additionaly to other nontrophic interactions (e. g., the raise of mycorrhizal networks in plant communities \cite{Heijden}, mutualism at weak direct competition \cite{Bastolla}, and facilitation \cite{Carrion}), the biochemical warfare between living organisms may drive species coexistence and community composition. The alternative view that biological communities can emerge from allelopathy, i.e., from competing interactions between their species mediated by toxins, faces a difficulty: multiple toxic environments are the least expected to sustain species diversity. Indeed, some exotic invasive plants may use allelopathic suppression to disrupt inherent, coevolved interactions among long-associated native species constituting the communities they invade \cite{Bais2,Souza}. Therefore, community and invasion ecology are naturally interconnected because both the persistence of a species in a community or its invasion success abroad its native habitat primarily depends on its ability to increase from low density \cite{Morton,Shea}.

In this report, our goal is to discuss how community structures of populations enforced to adapt and survive to the direct allelochemical suppression of each other is affected by the evolutionary history of the interaction. Specifically, we extend previously proposed models for the allelopathic warfare between two species \cite{Fassoni,Sylvestre} by integrating ecological and evolutionary processes. In the model, the genetic diversity is generated by mutations that induce changes in the allelochemical traits of the evolving species and selection is driven by ecological interactions, namely, intra- and interspecific resource competition and allelopathic suppression. These interactions determine how species evolve and enhance or diminish the diversity of communities. The paper is organized as follows. In Section \ref{model}, the mathematical model is introduced. The major results concerning community structure and biological diversity are reported in Section \ref{results}. In Section \ref{discussion} our major findings are discussed and some conclusions are drawn.

\label{model}
In order to model the community dynamics, a set $S$ of $l\in\mathbb{N}$ biological species with populations given by $\mathbf{N}=(N_1,N_2,\ldots)$ is considered. The interactions among these species occurs only via intra- and interspecific resource competition and allelopathic suppression. Thus, every species in $S$ synthesizes and releases toxic secondary chemical compounds (microcins, fitotoxins etc.) that enhance the mortality of other species. The strengths of such interactions depends on the toxin concentration $\mathbf{B}=(B_1,B_2,\ldots)$ and vary in time because $\mathbf{B}$ depends on the abundance of the species. Furthermore, the community assembly proceeds from an initial subset $S_0 \subseteq S$ by randomly adding new species through mutations fixed in a fraction of resident species' offspring.

\subsection{Ecological dynamics}
The temporal evolution of the biological community in a homogeneous environment is described by the coupled ordinary differential equations

\begin{eqnarray}
\label{pop_dyn}
 \frac{\mathrm{d}N_{i}}{\mathrm{d}t} &=& r_{i}\,\left( 1-\sum_{j= 1}^{l}\nu_{ij}N_{j}\right)\,N_{i} - \sum_{j\neq i}^{l}\mu_{ij}\Phi_{ij}^{(k)} (y_j) \,N_{i}\nonumber\\
 & & \\
 \frac{\mathrm{d}B_{i}}{\mathrm{d}t} &=& \beta_{i}\,N_{i} - \delta_{i}\,B_{i}- \sum_{j\neq i}^{l}\gamma_{ji}\,N_{j}\,B_{i}. \nonumber
\end{eqnarray}
Here, $N_{i}$ stands for the population density of the species $i$ that produces the allelochemical concentration $B_{i}$, respectively. Also, $r_i$, $\beta_i$ and $\delta_i$, $i = 1,2,\ldots$, respectively, are the reproduction rates, toxin release and natural degradation rates associated to the competing species. A classical interspecific competition for the environmental resources is assumed. The parameters $\nu_{ij}$ are the competition coefficients that measure the extent to which each species presses upon the resources used by the others. The quantity $y_j= \gamma_{ji} N_i B_j$ represents the overall consumption of the toxin $j$ by the species $i$, $i \neq j$, with per capta absorption rate $\gamma_{ij}$. These quantities depend on the toxin’s levels in a linear way. So, the term $-\sum_{j \neq i} \mu_{ij} \Phi_{ij}^{(k)}(y_j)$ represent species decreases as they uptake the allelochemicals released by their allelopathic suppressors, in which $\mu_{ij}$ is the mortality rate of the species $i$ induced by the toxin released by its competitor $j$. Different Holling type \textit{I}, \textit{II}, and \textit{III} functional responses were assumed:

\begin{eqnarray}
\label{res_func}
\Phi_{ij}^{(k)}=
\left\{\begin{matrix}
 B_{j} & (k=1) \\ \vspace{0.1cm}
 \gamma_{j,i}N_{i}B_{j} & (k=2) \\ \vspace{0.1cm}
 \frac{B_{j}}{c_{i}+B_{j}} & (k=3)\\ \vspace{0.1cm}
 \frac{\gamma_{j,i}N_{i}B_{j}}{c_{i}+\gamma_{j,i}N_{i}B_{j}} & (k=4)\\ \vspace{0.1cm}
\frac{B_{j}^{2}}{c_{i}+B_{j}^{2}} & (k=5)\\\vspace{0.1cm}
\frac{(\gamma_{j,i}N_{i}B_{j})^{2}}{c_{i}+(\gamma_{j,i}N_{i}B_{j})^{2}} & (k=6),
\end{matrix}\right.
\end{eqnarray}
where the parameters $c_i$ control the toxin’s efficiencies in poison their competing species. All these response functions assume null thresholds for toxin effects, but those with $k \geq 3$ impose saturation to the allelopathic suppression. Also, the response functions indexed by odd $k$'s involve the total toxin concentration, in contrast to those indexed by even $k$'s for which only the absorbed toxin can induce responses. 

Equations \ref{pop_dyn} and \ref{res_func} for two species were extensively investigated through analytical and numerical methods in references \cite{Fassoni,Sylvestre}. In the present paper up to $l=100$ competing species were considered and the interacting parameters $\nu_{i,j}$, $\gamma_{j,i}$, and $\mu_{i,j}$ define networks in which the species are the nodes. These parameters can be expressed as $\nu_{i,j}=\nu_{i,j}\,\varepsilon_{i,j}$, $\gamma_{j,i}=\gamma_{i,j}\,\zeta_{j,i}$, and $\mu_{i,j}=\mu_{i,j}\,\zeta_{i,j}$, in which $\varepsilon_{i,j}=1$ ($\zeta_{i,j}=1$) if species $i$ competes with (poisons) species $j$, but $\varepsilon_{i,j}=0$ ($\zeta_{i,j}=0$) if $i$ does not compete (poisons) $j$. Every $\varepsilon_{i,j}, \zeta_{i,j}=1$ is a link connecting two species.  The set of values $\varepsilon_{i,j}$ and $\zeta_{i,j}$ define two matrices $\varepsilon$ and $\zeta$ which characterize the competition and allelochemical interaction networks, respectively. These matrices are examples of the adjacency matrix, central in network theory \cite{Barabasi}. The diagonal elements of $\varepsilon$ are $\varepsilon_{i,i}=1$ and represent intraspecific competition, with all $\nu_{i,i}=1$ by definition. In turn, we set all $\zeta_{i,i}=0$ in order to avoid self-allelopathic suppression.

The ecological interactions (competition and allelopathy) drive the dynamics, equation \ref{pop_dyn}, towards an stationary state ($\mathbf{N}^*$, $\mathbf{B}^*$) in a short time scale. This stationary state depends on the species initially present and their interaction networks. Eventually, even in the weak interspecific competition (coexistence) regime, some populations are led to extinction by allelopathic suppression and the community diversity (species richness) decreases.

\subsection{Evolutionary dynamics}
The origin and maintenance of biological communities depends on the interplay between evolutionary processes and ecological interactions that allow species coexistence \cite{Edwards}. Ecological and evolutionary processes are integrated in our model by assuming that mutations in one of the competing species present at the current stationary state of the ecological dynamics generate a new species. This fresh species must survive and evolve in response to novel conditions, and the old species in the community must in turn evolve in response to the new species. Ultimately, the ecological dynamics is driven to another stationary state characterized by distinct populations and interaction networks. After that, additional genetic diversity is generated by adding different species to the current community, and so on. Two mechanisms for species introduction were tested.

\subsubsection{Sequential invasion events (SIE)}
An alien species, the node $n+1$, is added to an stationary state currently containing $n$ species. It is assumed that the alien species competes for resources with all the $n$ pre-existing species. Thus $\varepsilon_{n+1,i}=\varepsilon_{i,n+1}=1$ for $i=1,\ldots,n$. Concerning allelochemical suppression, the alien species affects $k_{n+1}^{out}$ of the old ones and is affected by $k_{n+1}^{in}$ of them. So, $k_{n+1}^{out}$ elements $\zeta_{n+1,i}$ in the line $n+1$ of the enlarged adjacency matrix $\zeta$ are fixed in $1$ and the remaining in $0$. In order to do this, an integer $i$ is randomly chosen in the interval $[1,n]$, and we set $\zeta_{n+1,i}=1$, with a probability $p=1-n^{out}/n$, or $\zeta_{n+1,i}=0$, with a probability $1-p=n^{out}/n$. Then, a distinct $i$ is randomly selected and the protocol repeated until $k_{n+1}^{out}$ elements in the $(n+1)$-th line of $\zeta$ are set to $1$. The value $n^{out} \in [1,n]$ defines the probability $p$ and, again, is an integer random number chosen with equal change. In average, $n^{out}$ determines the fraction of species in the community which do not interact with the alien species. Analogously, $k_{n+1}^{in}$ elements $\zeta_{i,n+1}$ in the column $n+1$ of the enlarged adjacency matrix are fixed in $1$ and the remaining in $0$. The same protocol is used to determine the $k_{n+1}^{in}$ nodes $i$ that suppress the node $n+1$ (i. e., $\zeta_{i,n+1}=1$). But now the probability used is $p=1-n^{in}/n$.  Finally, the initial toxin concentration of the alien species is $B_{n+1}=0$ and its population density is $N_{n+1}=0.01 N_i^*$, with $N_i^*$ corresponding to the stationary population density of one species chosen at random between the $n$ current members of the community. Regarding the initial community structure, the SIE evolutionary dynamics starts from a single species. 

\subsubsection{Branching process (BP)}
\begin{figure}
\centering
\includegraphics[width=9.5cm]{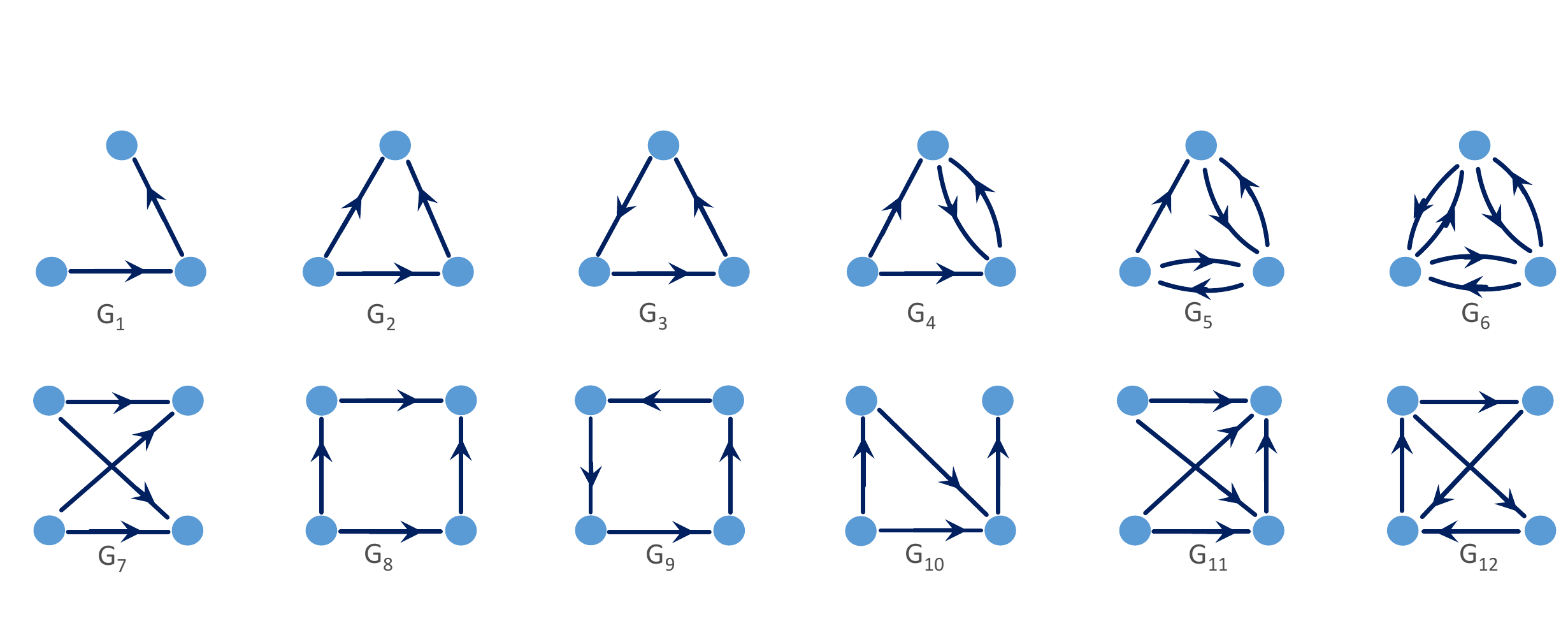} 
\caption{Allelopathic networks used as starting structures for the BP dynamics. The species interactions are indicated by arrows. In numerical integrations, the population densities $N_{i}(0)=0.7$ and toxin concentrations $B_{i}(0)=0$ were fixed.} 
\label{start_graphs}
\end{figure}

The new species $n+1$ introduced in the network descends from one of the $n$ species present at the community stationary state. The ancestor species $i$ is randomly chosen and only their allelochemical traits are mutated in its descendant species $n+1$. Specifically, all the $k_i^{in}$ input and $k_i^{out}$ output connections of the ancestor node $i$ are inherited by the new node $n+1$, except one of them. With equal chance, either a randomly chosen input $\zeta_{j,i}$ or output $\zeta_{i,j}$ of the node $i$ will be activated ($\zeta_{j,n+1}=1$) in node $n+1$ if inactive ($\zeta_{j,i}=0$) in $i$, or vice-versa. Since its is supposed here that the resource competition traits are not changed by mutations, $\varepsilon_{i,j}=\varepsilon_{n+1,j}$ and $\varepsilon_{j,i}=\varepsilon_{j,n+1}$ for $j=1,\ldots,n$. Again, the initial toxin concentration of the new species is $B_{n+1}=0$ and its population density is $N_{n+1}=0.01 N_i^*$. Finally, concerning the initial community structure, the BP evolutionary dynamics starts from a network with $n_0<l$ nodes. Different starting graphs for the BP dynamics are shown in figure \ref{start_graphs}.

\section{Community structures: species diversity and allelochemical network topologies}
\label{results}
The previously described eco-evolutionary process was investigated through numerical integration using the fourth-order Runge-Kutta method. Distinct distributions for the values of the competition and allelochemical parameters $\varepsilon_{ij}$ and $\zeta_{ij}$ were employed. Also, $200$ independent evolutionary histories were generated for the SIE and BP dynamics, in the latter case for each initial graph shown in figure \ref{start_graphs}. From the numerical integrations, the adjacency matrix at the successive stationary states for each evolutionary history were obtained. Then, the community structures (interaction network topology) and species richness were determined for both SIE and BP dynamics.

\subsection{SIE dynamics}
Since our primary interest relied on how allelopathic suppression affects the community structure, $\nu_{i,j}=\nu=0.1$ was fixed in order to ensure equal competition coefficients for every species in a regime of interspecific coexistence.

The scenario of equal (or homogeneous) allelopathic traits was investigated. Thus, each species has fixed toxin sensibility, $c_i=c=0.1$, release, degradation, and uptaken rates, $\beta_i=\beta=0.2$, $\delta_i=\delta=0.2$, and $\gamma_{j,i}=\gamma=0.1$, respectively, $ \forall i,j$. In turn, two mortality rates induced by allelochemicals were considered, namely, weak ($\mu_i=\mu=0.1$) and strong ($\mu_i=\mu=0.5$) $\forall i$.

\begin{figure}
\centering
\includegraphics[width=9cm]{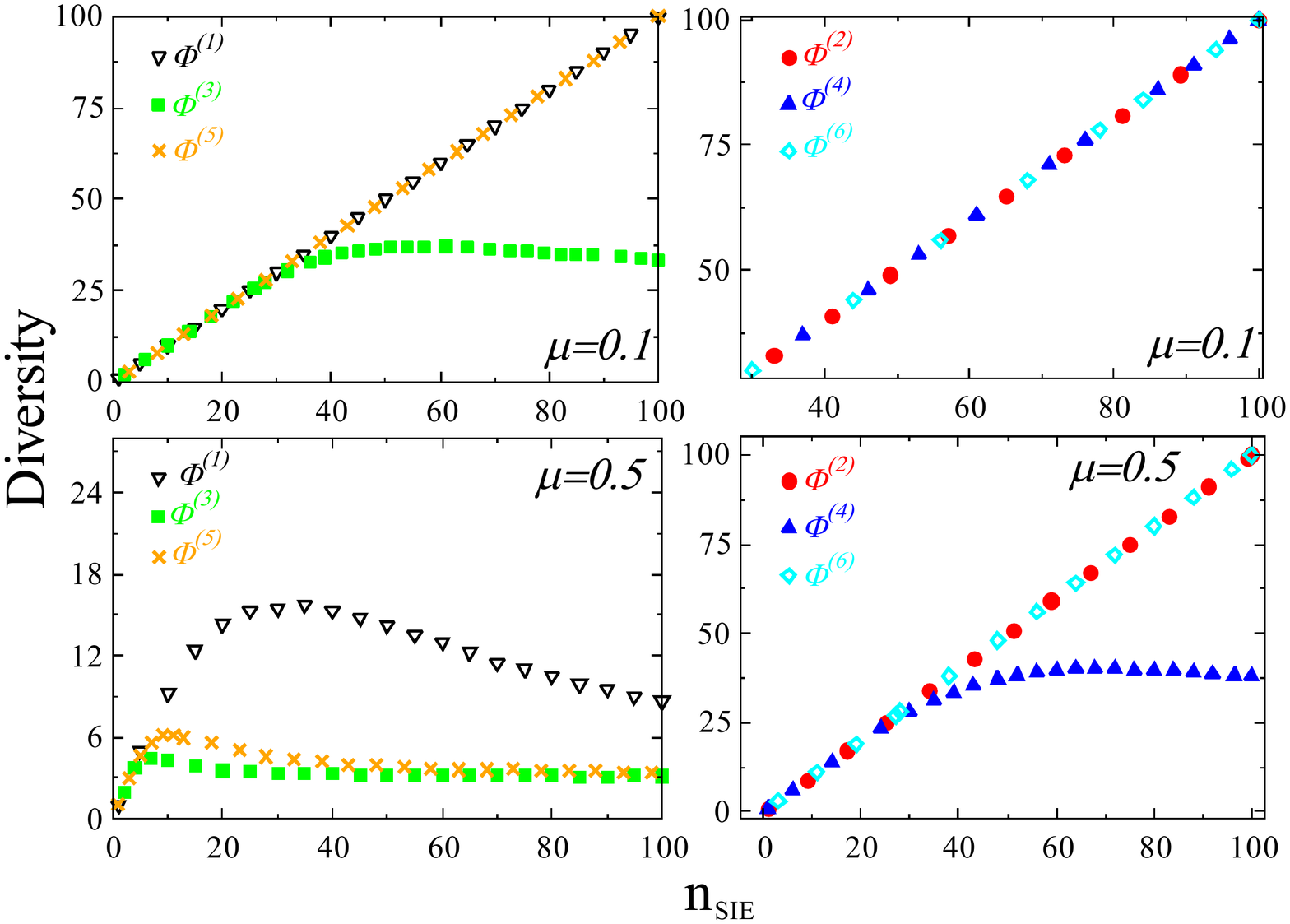} 
\caption{Average diversity for $200$ independent eco-evolutionary dynamics observed after successive invasion events. The initial community is always composed of a single species.  The top and bottom plots refer, respectively, to weak and strong allelopathic effects.}
\label{abundance_sie}
\end{figure}

In figure \ref{abundance_sie} it is shown the average diversity as a function of the number $n_{SIE}$ of SIE. The diversity or species richness is defined as the fraction of species that survive at the community stationary state. As expected, weak allelopathic suppression allows the assembly of communities exhibiting large diversities. This is true for all response functions tested and, as expected, the diversity decreases as the response to toxins increases. For instance, in our simulations, $\Phi^{(1)}(x) < \Phi^{(5)}(x) < \Phi^{(3)}(x)$ except for small ($x < 0.11$) or large ($x > 0.89$) toxin concentrations. In contrast, community diversity is drastically reduced at strong allelopathy for all response functions. As an example, the number of surviving species decreases from $\sim 100$, at weak, to $\sim 10$ at strong allelopathic suppression and response function $\Phi^{1}$. In this strong regime, diversity seems to decrease slowly after reaches a maximum as the number of invasion events increases. Also, the effect of toxin's uptaken is significant as revealed by the right column in figure \ref{abundance_sie}. In these graphs the response functions depend on the absorved fraction of toxins, not on their total concentration present in the homogeneous environment. So, even the regime of strong allelopathic suppression ($\mu=0.5$) at low toxins' absorption can become effectively equivalent to the weak ($\mu=0.1$) regime.

\begin{figure}
\centering
\includegraphics[width=9cm]{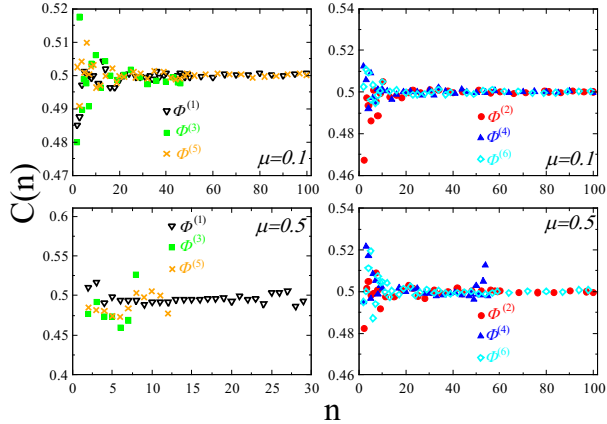} 
\caption{Average network connectivity $C(n)$ in communities containing $n$ species after a SEI. Again, the initial community is always composed of a single species. The results for weak and strong allelopathic suppression are shown in frames (a,c) and (b,d), respectively. The dashed lines corresponds to power laws with different exponents.}
\label{connect_sie}
\end{figure}

In figure \ref{connect_sie}, the average connectivity of allelochemical networks is illustrated as a function of the number $n$ of surviving species observed at the stationary state reached after a SIE. In a network of size $n$, the connectivity $C(n)$ is defined as the fraction of non-null elements in its $n \times n$ adjacency matrix ($\zeta_{i,j}$ in this case). In terms of the adjacency matrix $C(n)$ is given by 

\begin{equation}
C(n)=\frac{\sum_{i,j} \zeta_{i,j}}{n(n-1)}.
\end{equation}
Our results indicate that the average connectivity is significantly larger at weak ($\mu=0.1$) than strong ($\mu=0.5$) allelopathy. So, the interaction network is much more sparsely connected at strong allelochemical suppression. Furthermore, the connectivity initially increases up to $n \sim 10$ and saturates to a constant value and, for large response functions ($\Phi^{(3,5)}$), exhibits significant fluctuations at strong allelopathic regime. This behavior is very distinct from the power law scaling for large $n$ values observed in random networks, $C(n) \sim n^{-1}$ \cite{May}, and a model for growing random networks based on global stability, $C(n) \sim n^{-1.2}$ \cite{Perotti}. Therefore, our results indicate that the communities generated by the SIE dynamics markedly differs from random networks involving positive and negative interactions.

\begin{figure}
\centering
\includegraphics[width=9cm]{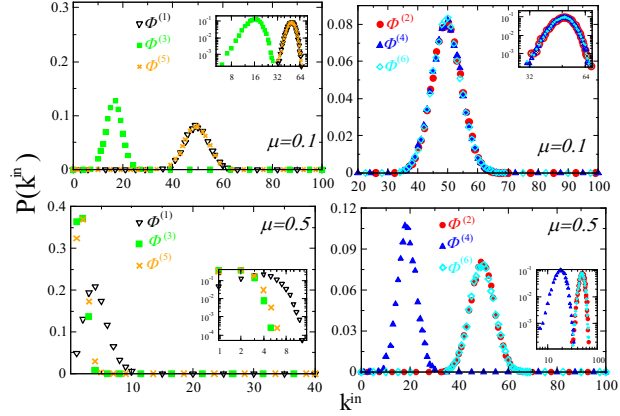} 
\caption{Degree distribution $P(k^{in})$ for SIE allelochemical interaction networks in which the competition and allelochemical traits are the same for all species. The top and bottom graphs correspond, respectively, to weak and strong allelopathic suppression.} 
\label{degree_pdf_sie}
\end{figure}

The degree distributions $P(k)$ for allelochemical interaction networks generated by the SIE dynamics are shown in figure \ref{degree_pdf_sie}. The distribution $P(k)$ gives the probability that a randomly selected node in a network has $k$ links, i. e., it is connected to $k$ nodes. Normal (Gaussian) and Weibull distributions was observed for in-degree distributions $P(k^{in})$ depending on the mortality $\mu$ and the functional response to allelopathy. For strong allelopathic suppression and functional responses $\Phi^{(1,3,5)}$, $P(k^{in})$ is a Weibull distribution. In contrast, at weak allelopathic suppression and for the response functions $\Phi^{(4,6)}$ at the strong regime, $P(k^{in})$ is Gaussian distributed. The apparent anisotropies observed in the insets for $\Phi^{(1,2,4,6)}$  are very weak, as supported by skewness $S \sim  0$ and kurtosis $K \sim 3$ (see the appendix). However, the ratio $\kappa=\langle k^2 \rangle / \langle k \rangle \sim \langle k \rangle$ is always obtained, indicating that the SIE allelochemical networks are homogeneous \cite{Barabasi}. In turn, the degree distributions $P(k^{out})$ for all scenarios are normal (Gaussian) distributions (see the appendix).

\begin{figure}
\centering
\includegraphics[width=9cm]{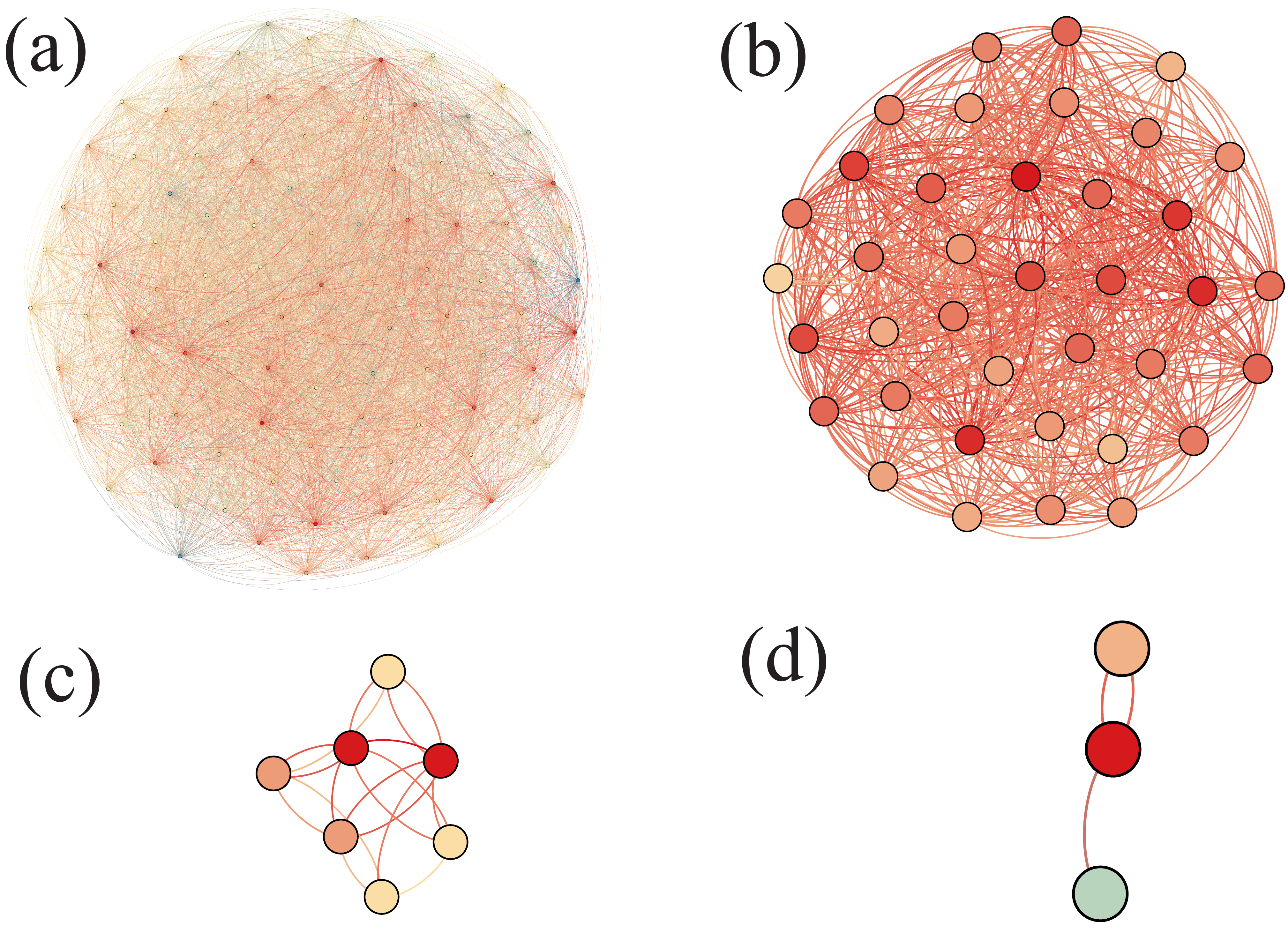} 
\caption{Typical allelochemical networks generated after $l=100$ SIEs for (a) weak ($\mu=0.1$) and (b)-(c) strong allelopathic suppression ($\mu=0.4$ and $\mu=0.5$, respectively). The competition and allelochemical traits are homogeneous (constant and equal for all species) and the functional response $\Phi_{i,j}^{(1)}$ was used.} 
\label{SIE_nets}
\end{figure}

Lastly, typical allelochemical networks or community structures generated by the SIE dynamics are illustrated in figure \ref{SIE_nets}. The nodes in these networks represent species present in the community and the directed edges between them represent allelopathic interactions. As the allelopathic strength increases, the number of node (surviving species) decreases, the network topology changes from random to hierarchical structures, and the corresponding connectivity distributions change from normal (or Gaussian) to Weibull distributions.

\begin{figure}
\includegraphics[width=9cm]{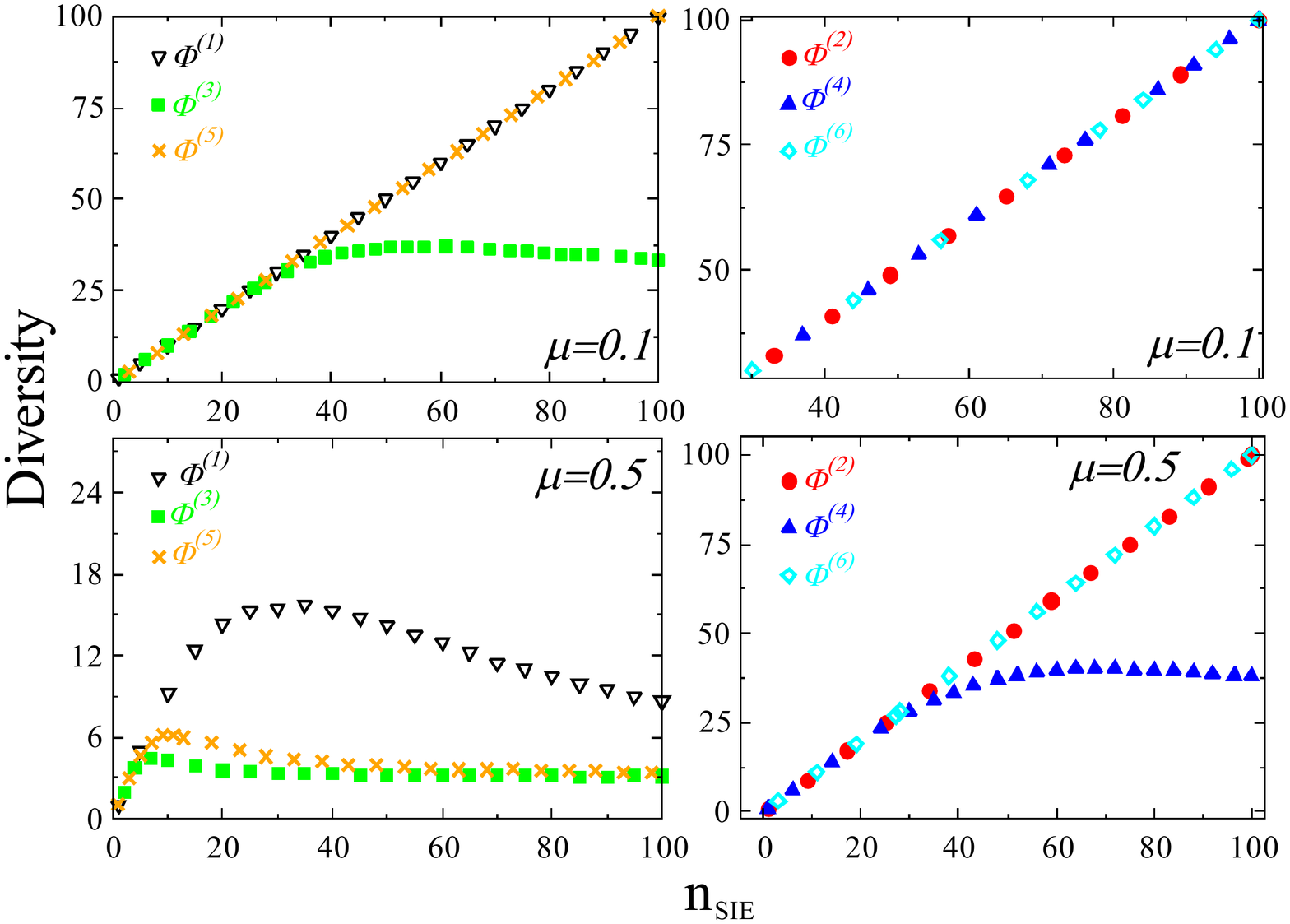} 
\caption{Average diversity for $200$ independent eco-evolutionary dynamics observed after successive invasion events at the strong allelopathic regime. The initial community is always composed of a single species.  In (a) both competition and allelochemical parameters are heterogeneous, i. e., randomly chosen from uniform distributions in $(0,1]$. In (b) only the competition coefficients are heterogeneous and the allelopathic traits are the same for all species. The value $\mu=0.1$ was used (weak allelopathy).}
\label{disorder_sie}
\end{figure}

The community structure was also investigated for scenarios in which only the competition or competition and allelopathy are heterogeneous, so that the coefficients $\nu_{i,j}$ and $\mu_{i,j}$ are drawn from random distributions. The results are qualitatively the same as those obtained for the homogeneous cases reported earlier (data not shown), but the introduction of heterogeneity in competition coefficients has smaller effects than in allelochemical parameters, as shown in figure \ref{disorder_sie}. Indeed, competition coefficients uniformly distributed in $[0,1]$ do not move the system from the competition coexistence regime, whereas allelochemical parameters, particularly, $\mu_i$ and $c_i$,  uniformly distributed in $[0,1]$ effectively correspond to strong allelopathy ($\langle \mu \rangle =0.5$).

\subsection{BP dynamics}
The BP dynamics was analysed for three distinct scenarios. In the first one, called homogeneous, all the original and introduced species have equal competition and allelopathic traits: $\nu_{i,j}=0.1$ and $\varepsilon_{i,j}=1, \, \forall \, i,j$, $\mu_{i,j}=0.1$ and $\gamma_{i,j}=0.1, \, \forall \, i \neq j$, $c_{i}=0.1$, $\beta_{i}=0.2$, and $\delta_{i}=0.2, \, \forall \, i$. In the second scenario, called heterogeneous competition, the allelochemical traits are equal, as before, but the competition coefficients $\nu_{i,j}$ are disordered, i. e., randomly drawn from a uniform distribution on the interval $(0,1]$. Thus, the species can have different competition, but the same allelochemical capabilities. Finally, in the third scenario, called completely heterogeneous, it is supposed that both competition and allelochemical traits are disordered and independently drawn from uniform distributions on the interval $(0,1]$. Only the toxins' degradation and uptaken rates, $\delta_{i}=\delta=0.3$ e $\gamma_{j,i}=\gamma=0.1$, are assumed the same for all species.

\begin{figure}
\centering
\includegraphics[width=9cm]{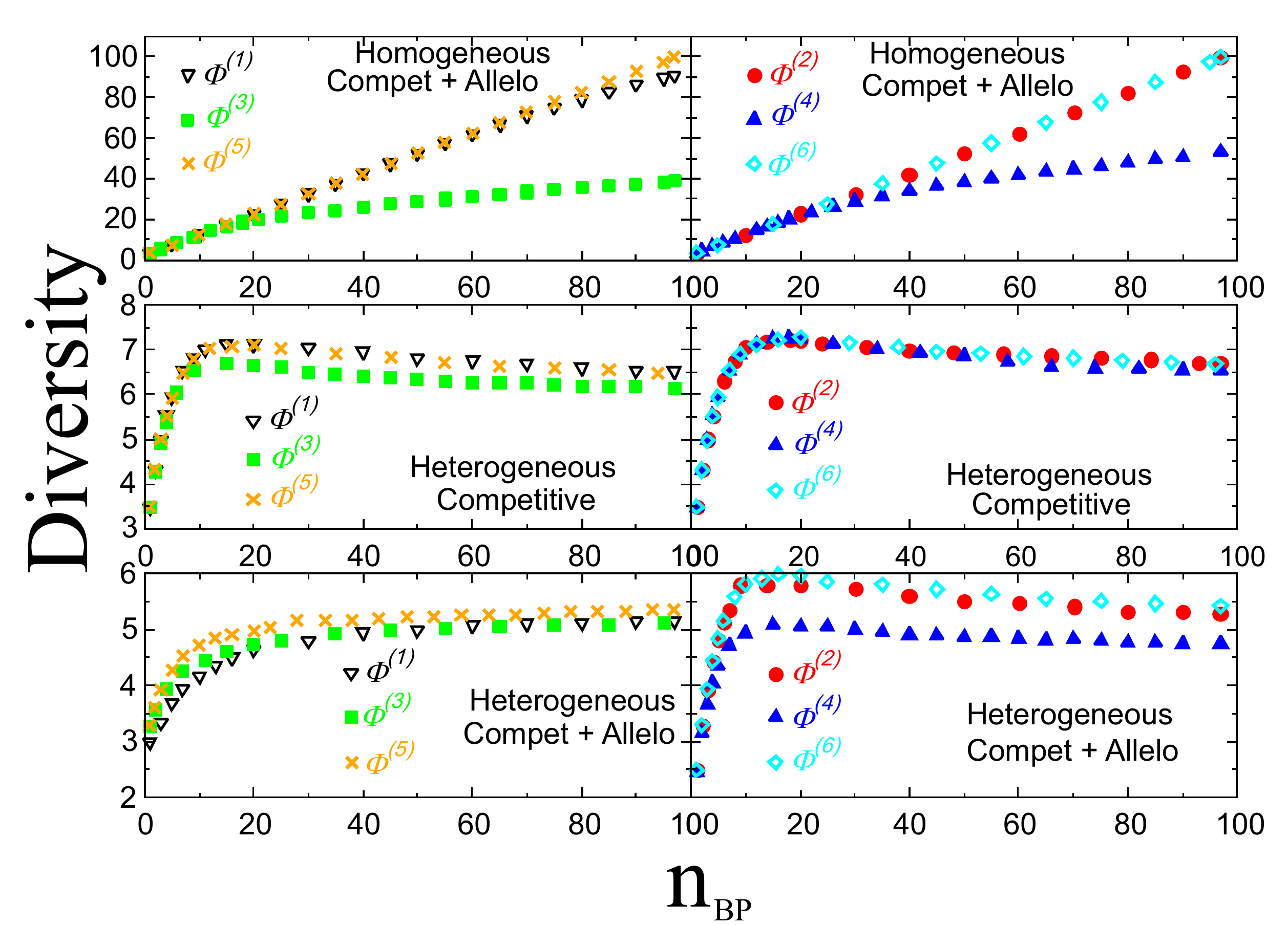} 
\caption{Average diversity for $200$ independent BP eco-evolutionary histories as function of the number $n$ of speciation events. The initial communities are the graphs shown in figure \ref{start_graphs}. The top, middle, and bottom plots refers, respectively, to homogeneous, heterogeneous competition, and completely heterogeneous (both competition and allelopathy) scenarios.} 
\label{abundance_bp}
\end{figure}

The average diversity as a function of the number $n$ of ``speciations'' is shown in figure \ref{abundance_bp}. Again, the response functions involving the total toxin concentration (odd $k$'s) induce more extinctions and lead to less diversity. Figures \ref{abundance_bp}b (disordered competition) and \ref{abundance_bp}c (disordered competition and allelopathy) evidence that heterogeneous competition and allelochemical traits decrease community diversity in comparison to homogeneous traits (figure \ref{abundance_bp}a).

\begin{figure}
\centering
\includegraphics[width=9cm]{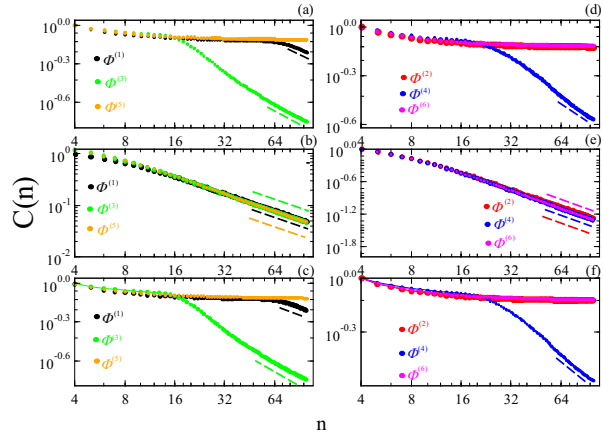} 
\caption{Average connectivity for communities grown from all graphs exhibited in the figure \ref{start_graphs} as a function of the network size $n$ for the homogeneous (top), heterogeneous competition (middle), and completely heterogeneous (bottom) scenarios. The dashed lines corresponds to the power law with different exponents.} 
\label{connect_bp}
\end{figure}

In figure \ref{connect_bp}, the average connectivity of allelochemical networks is illustrated as a function of the community size $n$. %\textbf{As for the SIE dynamics, the average connectivity ????????}.

\begin{figure}
\centering
\includegraphics[width=9cm]{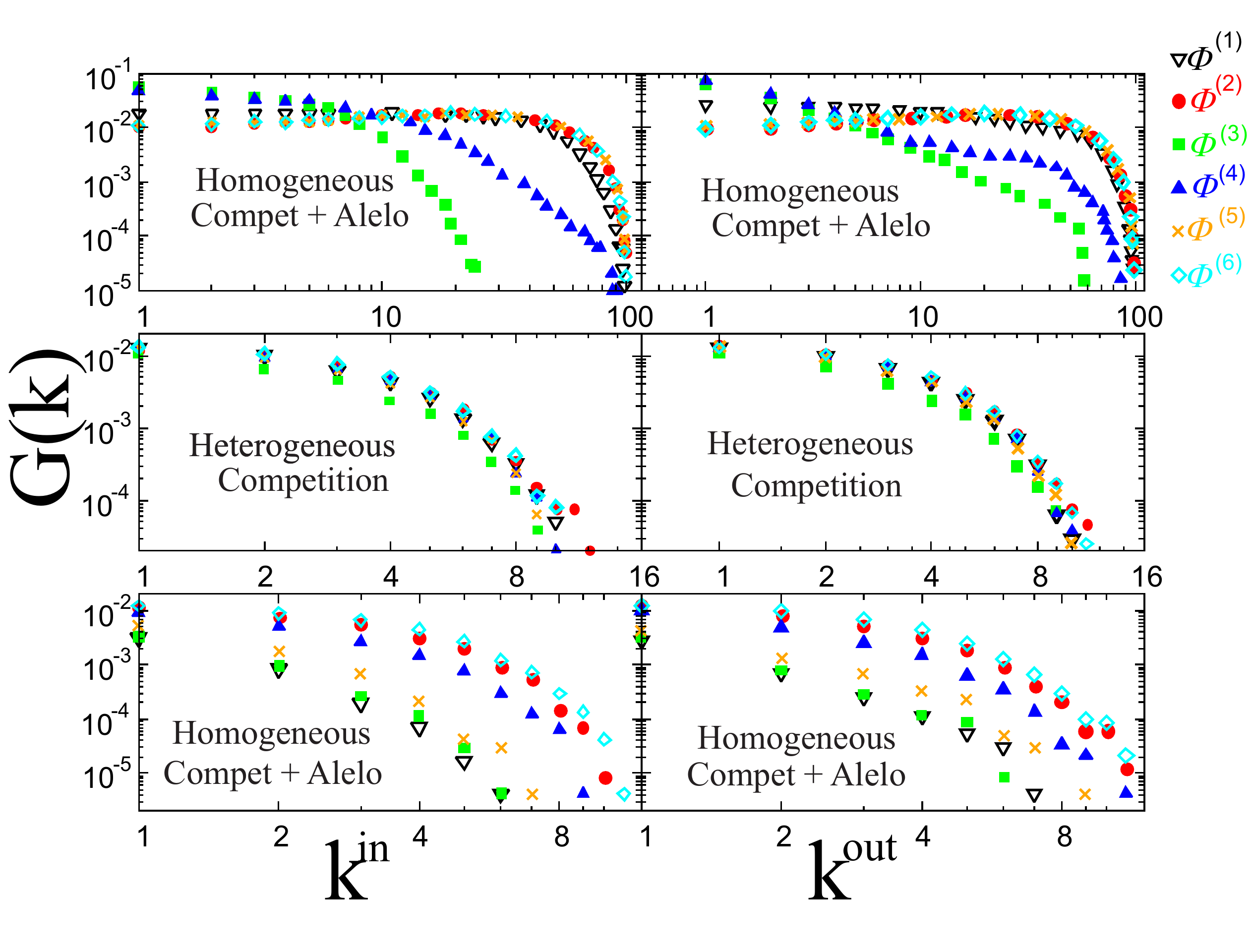} 
\caption{Complementary cumulative degree distribution functions for in- and out-degrees in directed allelochemical networks. The top, middle, and bottom plots refer, respectively, to homogeneous, heterogeneous competition, and completely heterogeneous scenarios.} 
\label{degree_pdf_bp}
\end{figure}

The in- and out-degree distributions, $P(k^{in})$ and $P(k^{out})$, for the allelochemical interaction networks generated by the BP dynamics are shown in figure \ref{degree_pdf_bp}. Indeed, the complementary cumulative degree distributions defined as 

\begin{eqnarray}
G(k^{in})  & =& 1-\sum_{k_i^{in} < k} P(k_i^{in}) \\ \nonumber
G(k^{out}) &= & 1-\sum_{k_i^{out} < k} P(k_i^{out}),
\end{eqnarray}
respectively, are ploted in figure \ref{degree_pdf_bp}. The cumulative distributions are used because they exhibit smaller statistical fluctuations than those observed for the degree distribution $P(k)$. In all the tested scenarios, these cumulative degree distributions are fitted by power-laws truncated by stretched exponentials $G(k) \sim k^{-\alpha} \exp(-\eta k^\lambda)$ (Weibull-like distributions). Again, $\langle k^2 \rangle / \langle k \rangle \sim \langle k \rangle$ is fulfilled, indicating the homogeneous nature of such networks. The values of this ratio and the exponents characterizing the degree distributions for SIE and BP dynamics are listed in the Appendix. 

\begin{figure}
\centering
\includegraphics[width=9cm]{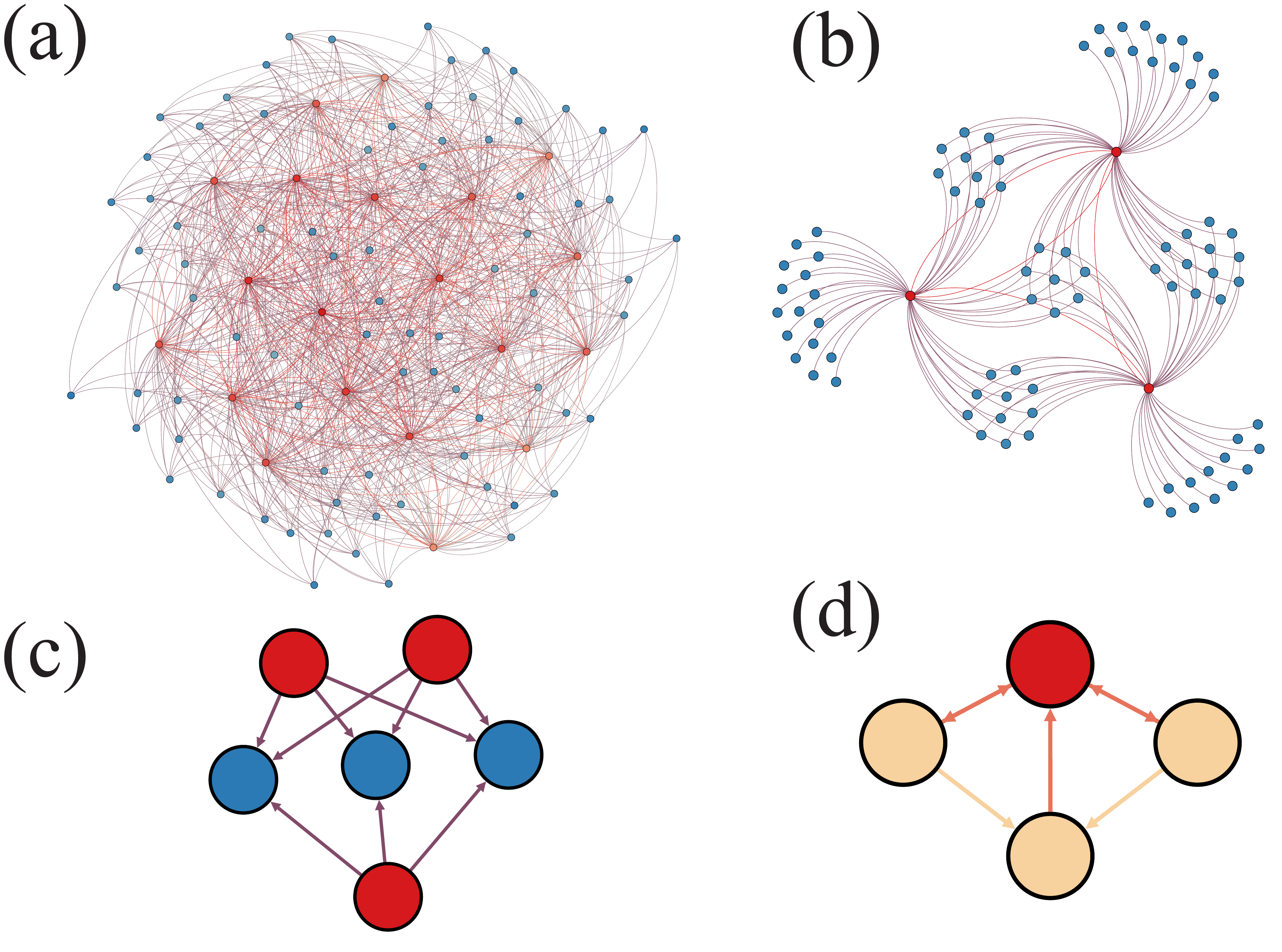} 
\caption{Examples of allelochemical networks generated after $l=100$ speciations for the homogeneous scenario (constant and equal competition and allelopathic parameters for all species) at weak allelopathy. The starting community structures are those shown in figure \ref{start_graphs}. The response functions $\Phi^{(3)}$ (left), $\Phi^{(4)}$ (top-right), and $\Phi^{(2)}$ (bottom-right) were used.}
\label{BP_nets}
\end{figure}

The BP dynamics generates largely diverse community structures as illustrated in figure \ref{BP_nets}. However, the hierarchical and modular character of such networks seems to be a universal trait. In order to further characterize these BP networks, the clustering coefficient, the average degree among the nearest neighbours of a node with degree $k$, the betweenness centrality and the network entropy were also determined.

\begin{figure}
\centering
\includegraphics[width=9cm]{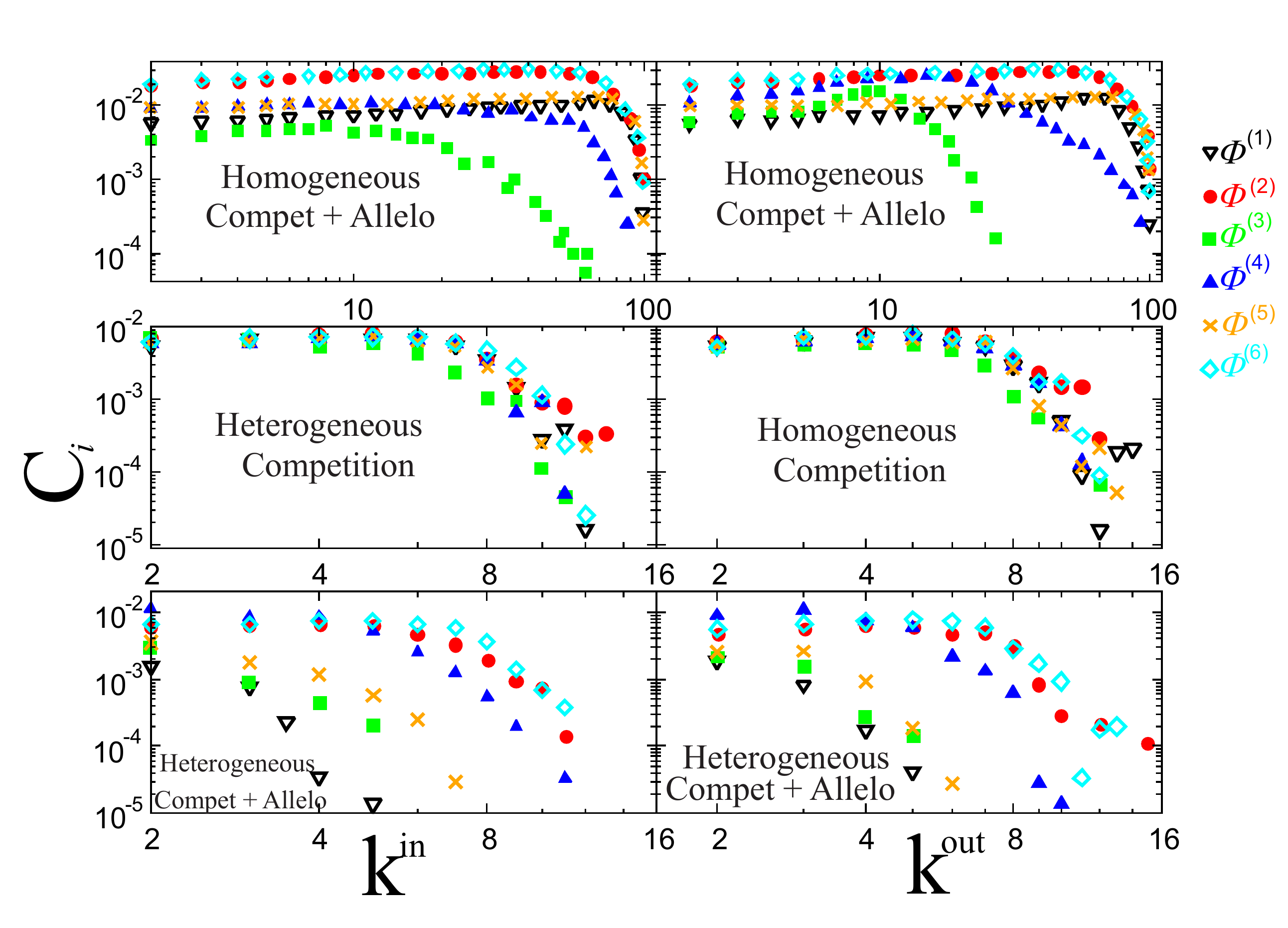} 
\caption{Average local clustering coefficient as a function of in- and out-degree. The top, middle, and bottom plots correspond to the homogeneous, heterogeneous competition, and completely heterogeneous scenarios. The initial networks were all those graphs shown in figure (\ref{start_graphs}).} 
\label{cluster_coef}
\end{figure}

The clustering coefficient $Cc$ measures the average probability that two nodes linked to another node are themselves linked to each other. In effect $Cc$ quantifies the density of triangles in a network \cite{Barabasi}. In the same way, we can define the local clustering coefficient $Cc_i$ for every node $i$, and a directed network has two such coefficients defined as

\begin{eqnarray}
Cc_i^{in}  & = & \frac{Tr(A^T A^2)}{k_i^{in}(k_i^{in}-1)} \\
Cc_i^{out} & = & \frac{Tr(A^2 A^T)}{k_i^{out}(k_i^{out}-1)},
\end{eqnarray}
where $A$ is the network adjacency matrix. Figure \ref{cluster_coef} shows the average local clustering as a function of the in- and out-degree. At weak allelopathic suppression (top and middle) and small responses, our results reveal that both $Cc^{in}(k)$ and $Cc^{out}(k)$ are constant for small degrees, but exhibit an exponential cut off for large degrees. However, despite the large fluctuations observed in $Cc_i^{in}$, the curves for the stronger response functions, $\Phi^{(3)}$ at the weak and $\Phi^{(1,3,5)}$ at the strong allelopathic suppression, suggest a power law scaling for large degree $k$. In turn, for $Cc^{out}$ this scaling is more neat. The exponents $b$ characterizing the power laws $Cc(k) \sim k^{-b}$ are close to one, a signature of modular structures with hierarchical organization \cite{Ravasz}. Since the behavior of Cc is associated to the dynamical mechanisms controlling which new attached node survives or extinguishes, this result indicates that allelochemical networks grow primarily by adding nodes with few links.

\begin{figure}
\centering
\includegraphics[width=9cm]{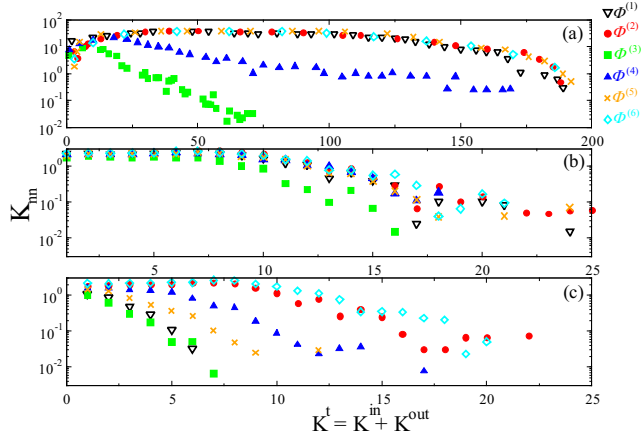} 
\caption{Average nearest neighbours degree $K_{nn}(k)$ of a node with total degree $k$ for the homogeneous (top), heterogeneous competition (middle), and completely heterogeneous (bottom) scenarios.} 
\label{nearest_neighbours}
\end{figure}

The average degree $K_{nn}$ among the nearest neighbours of a node with degree $k$ measures the mixing by degree properties of networks. $K_{nn}$ quantifies if there is a tendency of nodes with high degree to connect to others with high degree, and similarly for low degree. If this is the case, the network is assortative; if not, i. e., nodes with high degree tend to connect to others with low degree, the network is disassortative. In figure \ref{nearest_neighbours}, we see that $K_{nn}$ decays exponentially for $k \gtrsim 25$ at weak and $k \gtrsim 10$ at strong allelopathic suppression, in a clear disassortative behavior. Such a result is consistent with the observation that assortative mixing by degree makes a network more unstable \cite{Brede}.

\begin{figure}
\centering
\includegraphics[width=9cm]{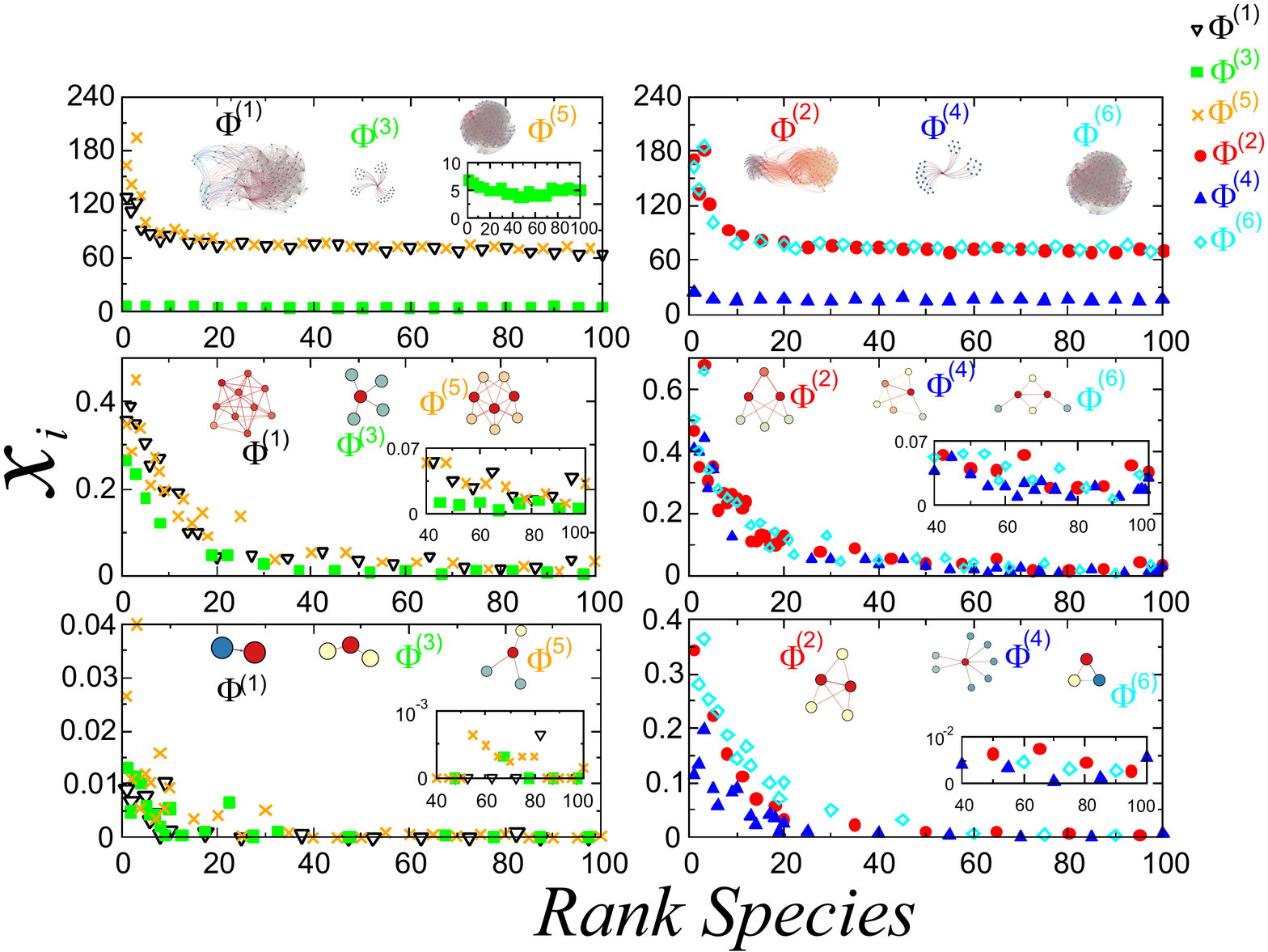} 
\caption{Average betweenness centrality for each node (surviving species) in communities generated from the initial graphs shown in figure \ref{start_graphs}. (a) homogeneous, (b) competition heterogeneous, and (c) completely heterogeneous scenarios.} 
\label{centralityBP}
\end{figure}

The betweenness centrality measures the extent to which a node lies on paths of minimal length connecting to other nodes \cite{Barabasi}. Nodes with high betweenness centrality often have significant influence on the network dynamics. Mathematically, the betweenness centrality $x_i$ of a node $i$ is defined as

\begin{equation}
x_i= \sum_{j,k} n^i_{jk},
\end{equation} 
where $n^i_{jk}=1$ if the node $i$ lies on the path of minimal length from node $j$ to node $k$ and $n^i_{jk}=0$ if $i$ does not or if there is no such path. In figure \ref{centralityBP}, the average $\langle x_i \rangle$ is plotted for every node $i$ present at the stationary allelochemical network after $l=100$ speciation events. It can be noticed that $\langle x_i \rangle$ decreases dramatically as the strength of allelopathic suppression increase. Indeed, even at weak suppression ($\mu=0.1$, figure\ref{centralityBP}(a)), strong responses to toxins ($\Phi^{(3,4)}$ lead to small or almost null average $\langle x_i \rangle$. Furthermore, the hierarchical and modular character of BP networks at weak allelopathy shown in figure \ref{BP_nets} is reflected on the peaks for small $k$ and the small fluctuations around a constant value of $\langle x_i \rangle$ for large $k$. In turn, a $\langle x_i \rangle =0$ for the heterogeneous competition and allelopathy is a consequence of very small and sparsely connected network structures. 

\begin{figure}
\centering
\includegraphics[width=9cm]{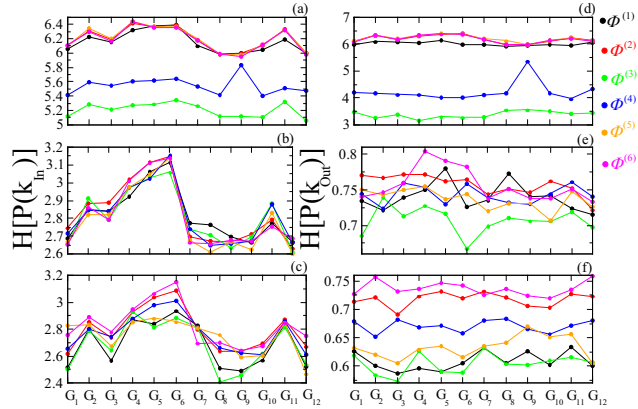} 
\caption{In- and out-degree distribution entropies for networks generated after $l=100$ speciation events starting from each initial community structures shown in figure \ref{start_graphs}. The upper, middle, and bottom plots correspond to the homogeneous, heterogeneous competition, and completely heterogeneous scenarios, respectively. Averages were taken over $200$ independent histories.} 
\label{entropy}
\end{figure}

Finally, the allelochemical network entropy was determined. Specifically, the degree distribution entropy $H$ for networks with size $n$, defined as 

\begin{equation}
H[P(k)]=-\sum_{k=k_{min}}^{k_{max}}P(k)\log_2(P(k)),
\end{equation}
was calculated. This Shannon entropy \cite{Shannon} is larger more homogeneous is the degree distribution and communities with greater diversities tend to be more homogeneous. As illustrated in figure \ref{entropy}, the entropy decreases as the allelochemical suppression increases. Particularly, $H[P(k^{out})]$ is more affected. Also, peaks at the initial community structures $G_2$, $G_6$, and $G_{11}$ are neatly observed in $H[P(k^{in})]$ for almost all response functions in the heterogeneous regimes. Peaks in the entropy $H[P(k^{out})]$ are less evident and restricted to few response functions, e. g., in $G_9$ for $\Phi^{(4)}$ and $G_4$ for $\Phi^{(6)}$ at the homogeneous and heterogeneous competition scenarios, respectively.

\section{Discussion}
\label{discussion}

We have proposed and studied, through numerical methods, an eco-evolutionary model for community assembly involving two coupled processes. The first is a fast ecological dynamics in which species compete for common resources and suppress each other allelopathically. The second are slow evolutionary events in which new species are added to the biological community at its ecological stationary state. Clearly, our study address a basic question: the relation between stability and complexity in the ecology of many interacting species. 

All the results obtained here must be analysed bearing in mind the scenario for pure intra- and interspecific competition. In the coexistence regime (weak competition, $\nu_{ij} < 1 \, \forall \, i,j$), all the surviving species at every stationary state constitute fully connected competition networks as assumed in our models. Since some introduced and/or resident species are eventually extinct, the community diversity tends to be smaller than the number of invasion or speciation events. Yet, communities with high diversity are the rule. This scenario changes if allelopathic interactions exist.

In the SIE dynamics, ecological networks grow through a succession of species imigration. These alien species allelochemically suppress and are suppressed by resident species at random, eventually leading to the eradication of either the invader or some resident species. Our results, shown in figures \ref{abundance_sie} and \ref{connect_sie}, reveal that communities exhibiting large diversities can be assembled at weak allelopathy, but diversities and average connectivities of stationary networks are drastically reduced at strong allelopathy for all response functions. Furthermore, in the strong suppression regime, species richness either saturates or decreases slowly after reaches a maximum. The maxima occurs after $\sim 10-30$ invasion events, depending on the response function to toxins (see figure \ref{disorder_sie} also). At the maxima, the average number of species in the communities never exceeds $16-18$. So, the system of interacting species becomes unstable and the networks stop to grow, consistent with the limit found by May \cite{May}. Beyond these upper bounds, the number of surviving species decreases continuously after each SIE until rest only one (a successful invasion) or very few species. Accordingly, network topologies evolve towards marked hierarchical structures, as seen in figure \ref{SIE_nets}, and the corresponding connectivity distributions change from normal (or Gaussian) to Weibull distributions (figure\ref{degree_pdf_sie}). Such networks, a subset of almost null measure in a random ensemble, can only be generated through a constrained growth process. 

The fundamental distinction between the SIE and BP dynamics is that new species are attached to the community with random or correlated connective patterns. %\textbf{?????}.

\section*{Acknowledgments}
This work was partially supported by the Brazilian Agencies CAPES (Carvalho graduate fellowship), CNPq (306024/2013-6 and 400412/2014-4), and FAPEMIG (APQ-04232-10 and APQ-02710-14).

\appendix

\begin{table}[]
\centering
\caption{My caption}
\label{my-label}
\begin{tabular}{|l|l|l|l|l|l|l|}
\hline
%                    & \multicolumn{2}{c|}{scenario1}              & \multicolumn{2}{c|}{scenario2} & \multicolumn{2}{c|}{scenario 3} \\ \hline
F R & $\frac{<k_{In}^{2}>}{<k_{In}>}$ & $\frac{<k_{out}^{2}>}{<k_{out}>}$ & $\frac{<k_{In}^{2}>}{<k_{In}>}$ &        $\frac{<k_{out}^{2}>}{<k_{out}>}$ & $\frac{<k_{In}^{2}>}{<k_{In}>}$ & $\frac{<k_{out}^{2}>}{<k_{out}>}$ \\ \hline
$\Phi^{(1)}$         & 44.2099 & 45.099  & 4.4801 & 47.5051 & 4.5526 & 54.7007 \\ \hline
$\Phi^{(2)}$         & 45.9834 & 45.5727 & 4.4781 & 47.8501 & 4.5162 & 48.7940 \\ \hline
$\Phi^{(3)}$         & 26.7295 & 47.8898 & 4.6144 & 49.3415 & 4.4764 & 54.8799 \\ \hline
$\Phi^{(4)}$         & 31.7409 & 45.643  & 4.4547 & 47.7538 & 4.5486 & 50.6062 \\ \hline
$\Phi^{(5)}$         & 46.2172 & 45.9558 & 4.1965 & 47.5544 & 4.6187 & 54.5946 \\ \hline
$\Phi^{(6)}$         & 46.1179 & 45.5262 & 4.4403 & 47.1904 & 4.5349 & 49.1369 \\ \hline
\end{tabular}
\end{table}

\end{document}